\documentclass[prb,twocolumn,showpacs]{revtex4}
\let\a=\alpha\let\b=\beta  
 \let\z=\zeta 
\let\th=\vartheta  \let\m=\mu 
   \let\s=\sigma \let\t=\tau
   
  \let\D=\Delta \let\Th=\Theta

\let\dpr=\partial\let\0=\noindent\let\fra=\frac

\def\V#1{{\,\underline#1\,}}
\def\media#1{\langle{#1}\rangle}

\def\Onlinecite#1{[\onlinecite{#1}]}

\newcommand\eg{{\it e.g.\ }}\newcommand\ie{{\it i.e.\ }}
\newcommand\dQ{{\not \kern-1.5pt d\, Q}}
\newcommand\dq{{\not \kern0.pt d\,Q}}
\newcommand\defi{\,{\buildrel def \over =}\,}
\newcommand\revtex{{R\kern-1mm\lower0.5mm\hbox{E}\kern-0.6mm V\kern-0.5mm%
\lower0.5mm\hbox{T}\kern-0.5mm E\kern-.5mm \lower0.5mm\hbox{X}}}

\newdimen\xshift \newdimen\xwidth \newdimen\yshift
\def\ins#1#2#3{\vbox to0pt{\kern-#2 \hbox{\kern#1
#3}\vss}\nointerlineskip}
\def\eqfig#1#2#3#4#5{ \par\xwidth=#1
\xshift=\hsize \advance\xshift by-\xwidth \divide\xshift by 2
\yshift=#2 \divide\yshift by 2 \hbox{\hglue\xshift \vbox to #2{\vfil #3
\includegraphics{#4.ps} }\hfill\raise\yshift\hbox{#5}}}
\def\8{\write13}

\begin{document}
\relax 

\preprint{RU/1-02g} 

\title{Nonequilibrium thermodynamics ?}

\author{G. Gallavotti} 
\affiliation{Rutgers Hill Center, I.N.F.N. Roma1, Fisica Roma1, I.H.E.S.}

\date{20 November 2002}
\begin{abstract}
Twentyseven comments $(\bullet)$ on the
Second Law and nonequilibrium systems
\end{abstract}

%\keywords{Nonequilibrium Thermodynamics, Entropy, Temperature}
\pacs{47.52, 05.45, 47.70, 05.70.L, 05.20, 03.20}

\maketitle

\begin{section}{\bf Definitions.}

The purpose, in the present discussion, is investigating the
possibility of an extension of Thermodynamics to systems which are in
a stationary state but are subject to the action of conservative and
nonconservative positional forces $\V f_{pos}$ and (therefore) also to
the action of the forces $\V \th$ necessary to take away the heat thus
generated.

We first consider systems for which a finite microscopic mechanical
model exists and are, therefore, described by equations of the form 
$$m\ddot{\V x}=\V f_{pos}(\V x)+\V\th(\V x,\dot{\V x})\defi \V F(\V
x,\dot{\V x})$$
and $\V x$ is a point in an appropriate finite dimensional phase space
(typically of very large dimension). If the force $\V f_{pos}$
is conservative then no thermostat is needed and we suppose
$\V\th=\V0$ (for simplicity). In general we call the force law $\V\th$
a {\it mechanical thermostat}.

\kern3mm\0($\bullet$) A key notion will be the {\it phase space contraction
rate} $\s(\dot{\V x},\V x)$ which is defined as minus the divergence
of the equations of motion: 

$$\s(\dot{\V x},\V x)=-\sum_{\a=1}^{3N}
\dpr_{\dot x_\a} F_\a(\dot{\V x},\V x)$$

\kern3mm\0($\bullet$) An equilibrium state will be a stationary
probability distribution given by a density (one says a stationary
``{\it absolutely continuous}'' distribution) on the phase space of a system
which is subject only to conservative forces. We also identify the
distribution with any point which is typical with respect to it: by
typical we mean that the time averages of observables evaluated on the
trajectory of the point are the same as the averages with respect to
the distribution.

\kern3mm\0($\bullet$) We suppose (ergodicity) that the time averages
of observables (just of the {\it few} physically relevant for
macroscopic Physics) are computable from any of the (equivalent)
statistical ensembles: like the microcanonical ensemble. Hence an
equilibrium state is identified with a probability distribution on
phase space.  A ``typical'' microscopic configuration, \ie an initial
datum in phase space which is not in a set of ``unlucky cases'' which,
however, form a zero volume set and are therefore (believed to
be)\footnotemark[1]\footnotetext[1]{This is not to be taken for
granted, even though it is very often considered so: here I do not
enter into discussing this mysterious assumption (as I have nothing to
say).} unobservable, will evolve in time so that the time averages of
the observables (at least the {\it few} relevant for macroscopic
Physics) are computable by the equilibrium state which has the correct
values of the macroscopic parameters: \eg in the microcanonical case
the energy $U$ and the container volume $V$. When we speak of
properties of a single (typical) point in phase space, like of its
``entropy'', we always mean the same property of the equilibrium state
for which the datum is a typical one.

\kern3mm \0{\it Remarks:} This already might be controversial: in fact
the above (admittedly unconventional) definition of entropy has the
following implications. A rarefied gas which initially happens to have
all molecules located in the left half of a container, because a
separation wall has just been removed, setting the gas in macroscopic
motion and out of the previous equilibrium state, will be an initial
datum in phase space whose entropy is that of the same gas occupying
the entire container and at the same
temperature\footnotemark[2]\footnotetext[2]{Assuming the gas to be
ideal: we think here of Joule's experiment. In the following the
temperature will have the dimension of energy, \ie we call temperature
$T$ what is usually called $k_B T$ with $k_B$ being Boltzmann's
constant} (the difference residing only in the different dynamics that
follows the removal the wall in the middle of the container). Since a
physicist would apply the Boltzmann equation to describe the
evolution, the question arises about which is the place here of
Boltzmann's $H$ function, which is different if evaluated for the
initial datum or for a datum into which the initial one evolves after
a moderately large time (and in both cases it equals the classical
thermodynamic entropy of the initial and final equilibria).  \*
    
\kern3mm\0($\bullet$) In trying to study nonequilibrium cases a
conceptual difficulty must be met: if a system is subject to external
non conservative forces then the thermostatting forces will have a non
zero divergence and volume in phase space will not be preserved. The
key idea (due to Ruelle)\Onlinecite{Ru95} is that in this case (as well as
in the previous case) the nonequilibrium states will be the stationary
states which are generated by time averaging of initial states that
have a density or more satisfactorily, perhaps, by time averaging on
the evolution of initial data that are typical for probability
distributions given by some (arbitrary) density. The latter states are
called {\it SRB distributions}
(Sinai,Ruelle,Bowen)\Onlinecite{Ru95,Ru99,Ga00,Ga02}. Their appearance is
natural since there will be {\it no state} (\ie no probability
distribution) which is stationary and at the same time is also given
by a density.  As in the equilibrium state we shall attribute to each
individual point in phase space the macroscopic properties of the
stationary state which allows us to compute the averages of
macroscopic observables on the motion of the given point.

\kern3mm \0{\it Remark:} Systems that are chaotic in a mathematical
sense ({\it hyperbolic systems}) can be shown, on rather general
grounds, to have the property that there is only one SRB distribution
(with the correct values of the macroscopic
parameters).\Onlinecite{Ru95,Ru99,Ga02} Therefore such systems, the only
ones for which the above definition has a strict mathematical
sharpness, verify an extension of the ergodic hypothesis: adopting the
latter definition means believing that the system is {\it chaotic
enough} so that typical initial data generate a unique stationary
distribution with several features of the SRB distributions for
hyperbolic systems: the assumption has been called {\it chaotic
hypothesis}\Onlinecite{GC95} and it represents (in our analysis) the nonequilibrium
analogue of the classical ergodic hypothesis.  \kern3mm

\kern3mm\0($\bullet$) Given an initial state (a typical point or a distribution
on phase space) it might be possible to define a function of it that
will monotonically increase to a limit value which is the same for
almost all data sampled with a distribution with a density on phase
space (\ie absolutely continuous). Since Boltzmann's $H$ function is
an example of such a function we shall call $H$-functions all such
functions. We recall that Boltzmann's $H$ function is defined by a
coarse graining of phase space into ``macrostates'' determined by the
occupation numbers $f(p,q)d^3pd^3q$ of phase space cells $d^3p d^3q$
around $p,q$ and by defining $H=-\int f(p,q)\log f(p,q)\,d^3pd^3q$.
Clearly here the cells size affects by an additive constant the actual
value of $H$, which therefore should have no significance (at least in
the classical mechanics context in which we are working) and only the
variations of $H$ can be meaningful.  In a rarefied gas the Boltzmann
equation is believed to apply:\Onlinecite{Ga00} so that in the latter cases the
function just defined is a nontrivial example of an
$H$--function.

\kern3mm\0($\bullet$) Here we propose that even in the case of rarefied gases it
is neither necessary nor useful that the $H$ function is {\it
identified with the entropy}: we want to consider it as a Lyapunov
function whose role is to indicate which will be the final equilibrium
state of an initial datum in phase space. This does not change, nor it
affects, the importance of Boltzmann's discovery that, in rarefied
gases, the $H$ function can be identified with the physical entropy
whenever the latter is defined (\ie in equilibria). Suppose that the initial
state is chosen randomly with respect to a Gibbs distribution which is
{\it not} the one that pertains to the given parameters that describe the
system (\eg volume $V$ and energy $U$) but to other values: for
instance it is chosen randomly with a Gibbs distribution $\m_{U,V/2}$
that has the same energy but occupies half of the volume as in Joule's
experiment. Then the inital and final values of the $H$ function
happen to coincide with the physical entropies of the Gibbs states
$\m_{U,V}$ and $\m_{U,V/2}$ (at least in a rarefied
gas) which are given by the Gibbs' entropy.\footnotemark[3]\footnotetext[3]
{One should note that in the whole Boltzmann's work he has been really
concerned with the approach to equilibrium: in our terminology he has
been concerned with the problem of determining the stationary
equilibrium distribution to which a given initial datum gives rise at
large time: restricting consideration only to such cases one could
well call the Boltzmann's $H$-function the ``entropy'' of the state as
it evolves towards equilibrium. And, whatever name we give it, remains
true that $H$ is a measure of the disorder in the system. Our analysis
here is intended to say that such an interpretation is not tenable
when the system evolves towards a stationary nonequilibrium state. It
must also to be said that even the H-theorem is {\it not general}
because it applies only to rarefied gases, and even there it is an
approximation: to generalize it to general situations, even when one
only deals with approach to equilibrium, is a profound statement which
should be substantiated by appropriate arguments.}

\kern3mm\0($\bullet$) The latter property could possibly be used to
attempt a definition of entropy for states which are neither
equilibrium nor stationary states:\Onlinecite{Le93} however such a
definition would be useful only in the special situations in which an
H-theorem could be proved. That seems effectively to reduce the cases
in which the notion would be useful to the ones in which an initial
equilibrium state identified by some parameters (like $U,V$) evolves
towards a final one identified by other values of the parameters. And
even in such cases it is severely restricted to the rarefied gases
evolutions in which the H-theorem can be proved: a statement that a
model indpendent, universal, extension of the above $H$-function,
would have to be proved to have a monotonicity property, at least at a
heuristic level, to avoid that its monotonicity becomes an {\it a
priori} law of nature.

\kern3mm\0($\bullet$) In general I would think that there will always
be a Lyapunov function which describes the evolution of an initial
state and is maximal on the stationary state that its evolution will
eventually reach: however such a Lyapunov function may not have a
universal form (unlike the $H$ function in the rarefied gases cases)
and it may depend on the particular way the system is driven by the
external forces. After all the SRB distribution verifies a variational
principle (Ruelle)\Onlinecite{Ru95,Ru99,Ga02} which {\it remarkably} has the
same form both in equilibrium and nonequilirium systems and one may
imagine that in general it will be possible to define (on a case by
case basis, I am afraid) a quantity that will tend to a maximum
reached on the eventual stationary state. This picture seems to me
simpler than trying to guess a (possibly nonexistent) general
definition of a quantity that would play the role of Boltzmann's $H$.
\end{section}

\begin{section} {Entropy creation.}

The second law of Thermodynamics, in classical Thermodynamic
treatises, states:

\kern3mm
\0{\it It is impossible to construct a device that, operating in a
cycle, will produce no effect other than the transfer of heat from a
cooler to a hotter body}

\kern3mm\0Of 
course this will be assumed to be a law of nature (Clausius)\Onlinecite{Ze68}.

\kern3mm\0($\bullet$) The law implies that one can define an {\it entropy}
function $S$ on all equilibrium states of a given system
(characterized in simple bodies by energy $U$ and available volume
$V$) and if an equilibrium state $1$ can be transformed into another
equilibrium state $2$ then

$$S_2-S_1\ge \int_1^2 \fra{\not \kern-1.5pt d\, Q}{T}$$
using notations familiar from Thermodynamics: here the integral is
over the transformation followed by the system in going from $1$ to
$2$ and the $\not \kern-1.5pt d\, Q$ is the heat that the system
absorbs at temperature $T$ from the outside reservoirs with which it
happens to be in contact. The equality sign holds if the path followed
is a reversible one.

\kern3mm\0($\bullet$) One should note that the principle really says
that the $\int_1^2\fra{\dq}T$ does not depend on the path followed, if
the path is a reversible sequence of equilibrium states; and its
maximum value is reached along such paths. Existence of a path
connecting $1$ with $2$ with $S_2-S_1<\int_1^2 \fra{\dq}{T}$ would
lead to a violation of the second law.

\kern3mm\0($\bullet$) Is there an extension of the $S_2-S_1\ge
\int_1^2 \fra{\dq}{T}$ relation to nonequilibrium
Thermodynamics?\footnotemark[4]\footnotetext[4]{Which in a sense is
tantamount of asking ``is there a nonequilibrium Thermodynamics?''}
Since there seems to be no agreement on the definition of entropy of a
system which is in a stationary nonequilibrium and since there seems
to be no necessity in Physics of such a notion, at least I see none, I
shall {\it not} define entropy of stationary nonequilibrium systems
(in fact the analysis that follows indicates that if one really
insisted in defining it then its natural value could, perhaps, be
$-\infty$!).

\kern3mm\0($\bullet$) In Thermodynamics one interprets $-\int_1^2\fra{\dq}{T}$
as the {\it entropy creation} in the process leading from $1$ to $2$,
{\it be it reversible or not or through intermediate stationary states
or not}. The name is chosen because it is thought as the entropy
increase of the heat reservoirs with which the system is in contact
and which are supposed to be systems in thermal equilibrium: so that
their entropy variations are, in principle, well defined because they
fall in the domain of equilibrium thermodynamics. In a general
transformation from a state $1$ to a state $2$, both of which are
stationary (non)equilibrium states, following a path of
(non)equilibrium stationary states $\m^{(t)}$ and in contact with
purely mechanical thermostats one could consider the contribution to
the {\it entropy creation} due to irreversibility in the process
leading from $1$ to $2$ during a time interval $[0,\Th]$ to be

$$\D=c \int_0^\Th \media{\s(\V x,\dot{\V x})}_{\m^{(t)}} dt=c\int_0^\Th
\s_t\, dt$$
where $\s_t\defi\media{\s(\V x,\dot{\V x})}_{\m^{(t)}}$ is the average
phase space contraction computed in the state $\m^{(t)}$. This follows
a recently proposed identification of $\s(\V x,\dot{\V x})$ as
proportional to the {\it entropy creation rate} (here $c$ is a
proportionality constant)\Onlinecite{EM90}. The quantity $\D$ is for mechanical
thermostats the analogue of $-\int_1^2\fra{\dq}{T}$ for the generic
phenomenological thermostats characterized by a temperature $T$.

\kern3mm\0($\bullet$) In considering macroscopic systems one may imagine
situations in which a system is partially thermostatted by mechanical
forces for which a model considered physically reasonable is
available\footnotemark[5]\footnotetext[5]{Typically these are models of
friction, as in the Navier Stokes equation case in which the viscosity
plays the role of a thermostat. Or in granular matter where the
restitution coefficient in the collisions produces energy
dissipation. Another well known example is in Drude's theory of
electrical conduction.} and partially by phenomenologically
defined ``heat reservoirs'' characterized by a temperature $T$ and
able to cede to the system quantities $\dQ$ of heat ($\dQ$ can have
either sign or vanish).

In such more general settings a system in contact with several
thermostats of which a few are modeled by mechanical equations and a
few others are unspecified and are just assumed to exchange quantities
of heat $\dQ$ at temperature $T$ the second principle will be extended
as

$$-\D+\int_1^2 \fra{\not \kern-1.5pt d\, Q}{T}\le0$$
{\it assuming that the path leading from $1$ to $2$ consists entirely
of nonequilibrium stationary states} and that it lasts a time interval
$[0,\Th]$.  Regarding the external thermostats as thermodynamic
equilibrium systems $-\fra{\dq}T$ is the entropy increase of the
reservoirs at temperature $T$ and $\D= c \int \media{\s}_t$ is
interpreted as the entropy increase of the mechanical reservoir: if
this interpretaton is accepted the above relation becomes the ordinary
second law for the external reservoirs and could be read as ``the
entropy of the rest of the universe does not decrease'' (because
$\D-\int_1^2 \fra{\dq}{T}\ge0$), where ``universe'' is not the
astronomical Universe but rather the collection of physical systems
whose interaction with the system under study cannot be neglected.

If a system is a in a stationary state in which $\s_t=\media{\s}>0$
($t$--independent) this essentially forces us to say that its entropy,
if one insisted in defining it at the time $t_0$ of observation, could
only be $-\int_{-\infty}^{t_0}\media{\s}\,dt=-\infty$ as hinted
above.\footnotemark[6]\footnotetext[6]{If we imagine possible to
replace a mechanical thermostat with a phenomenological thermostat at
temperature $T$ then the left hand side of the relation above remains
unchanged but a part of $-\D$ becomes a contribution to the second
addend.}

\kern3mm\0($\bullet$) Since the quantity $\s_t$ is $\ge0$ 
(Ruelle)\Onlinecite{Ru99}
under very general conditions (in fact always if the chaotic
hypothesis is assumed to hold true) $\D\ge0$ and the proposed
estension is compatible with the main consequences of the second
law. The constant $c$ will be taken $1$ because the factor $1$ can be
computed by studying the expression of $\s(\V x,\dot{\V x})$ in
special models: at the moment, however, I see no immediate physical
implications of the ``universality'' of this choice of $c$ and for the
purposes of what follows $c$ could be any constant, even non
universal.

\kern3mm\0($\bullet$) The above implies again that we shall {\it not} be
able to define an entropy function {\it unless} $\s_t\equiv0$. The
latter is the condition under which equilibrium Thermodynamics is set
up: so that if one studies only transformations from equilibrium
states to other equilibrium states it is possible to define not only
the creation of entropy but the entropy itself (up to an additive
constant).

\kern3mm\0($\bullet$) A number of compatibility questions arise:
suppose that the system evolves between $1$ and $2$ under the action
of a thermostat which is modeled by forces that act on the system. For
instance we can imagine a container with periodic bounday conditions,
we call it a {\it wire}, containing a lattice of obstacles, which we
call a {\it crystal}, and $N$ particles, which we call {\it
electrons}, interacting between each other and with the lattice via
hard core interactions (say) and subject also to a constant force,
which we call {\it electromotive force}, of intensity $E$; furthermore
the particles will be subject to a thermostat force of Gaussian
type\footnotemark[7]\footnotetext[7]{This is not the appropriate place
to remind the Gauss' least constraint principle: it can be easily
found in the literature\Onlinecite{Ga00}.} (as it is essentially the case in
Drude's theory as described in classical electromagnetism
treatises\Onlinecite{Be64}) which forces the particles to have an energy
$U=u(E)$ which is an assigned function of $E$. Suppose that the value
of $E$ changes in time (very slowly compared to the microscopic time
scales) following a profile $E(t)$ as drawn in Fig.1

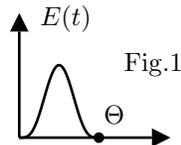
\eqfig{100pt}{50pt}
{\ins{8pt}{50pt}{$E(t)$}
\ins{32pt}{12pt}{$\Th$}}
{fig1}{Fig.1}
\0In this case the wire performs a cycle which is irreversible and the
integral $\int_1^2 \fra{\not \kern1pt d\, Q}{T}$ is $0$ because the
system is adiabatically isolated (the thermostat being only of
mechanical nature). The entropy variation of the system is defined
because the initial and final state are equilibria and, since they are
the same, it is $0$: but there has been entropy creation $\D>0$.

\kern3mm\0($\bullet$) {\it It is} (perhaps) {\it natural to define the
``temperature'' of a mechanical thermostat} by remarking that in the
models studied in the literature it turns out that $\s_t$ is
proportional to the work per unit time that the mechanical forces
perform, the porportionality constant being in general a function of
the point in phase space. Therefore we can call $T_0^{-1}$ the time
average of the proportionality constant between $\s$ and the work $W$
per unit time that the mechanical thermostatting forces perform: in
this way $\s_t=\fra{W}{T_0}$ and $\int_0^\Th \s_t\,dt=\int
\fra{\dq_0}{T_0}$ where $\dQ_0$ is the total work performed by the
mechanical forces {\it which we can} (naturally) {\it call the heat
absorbed by the mechanical reservoir}.

\kern3mm\0($\bullet$) If one imagines that the above conducting wire model at
the same time exchanges heat with two sources, absorbing
$Q_2$ at temperature $T_2$ and ceding $Q_1$ at temperature $T_1$ via
some unspecified mechanism, and assuming that the profile of $E(t)$ is
as in Fig.2

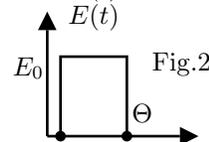
\eqfig{100pt}{50pt}
{\ins{8pt}{50pt}{$E(t)$}
\ins{-13pt}{30pt}{$E_0$}
\ins{32pt}{12pt}{$\Th$}}
{fig2}{Fig.2}

\0where the value of $E_0$ corresponds to a temperature $T_0$ in the
above sense. The inequality $-\D+\int_1^2 \fra{\dq}{T}\le0$ becomes

$$-\int_0^\Th \s_t\,dt\, +\, \fra{Q_2}{T_2}-
\fra{Q_1}{T_1}\equiv
-\fra{Q_0}{T_0}+\, \fra{Q_2}{T_2}- 
\fra{Q_1}{T_1}\le0$$
For instance, we see that if $T_1=T_2=T_0$ we have realized a cycle
which is irreversible. In it a quantity of heat $Q_2-Q_1-L$, with
$L=Q_0= T_0\D=T_0 \int_0^\Th \s_t\,dt$ is absorbed at a single
temperature $T=T_0$ and is transformed into the amount $Q_2-Q_1-L$ of
work: however the inequality forbids this to be positive, as expected.

\end{section}

\begin{section}{Mechanical and stochastic models.}

A definition in Physics is interesting (only) if it is useful to decribe
properties of the systems in which we are interested. Therefore having
set the above definitions one should expect to be asked why all the
work was made.

In this case the whole matter was originated by efforts to interpret
results that started to appear in the late 1970's concerning numerical
experiments in molecular dynamics\Onlinecite{EM90}.

\kern3mm\0($\bullet$) It is obvious that in numerical experiments one needs to
deal with a finite system (and even not too large): hence various
models of thermostats were devised for the purpose of obtaining
equations that could be transformed into numerical codes and studied
on electronic machines. This was a theoretical innovation with respect
to previous models which either relied on stochastic boundary
conditions or, in the more sophisticated cases, with (poorly
understood) systems with infinitely many particles. And it opened the
way to import the knowledge in the theory of dynamical systems that
had been being developed in the two preceding decades or so.

The novelty with respect to stochastic thermostats was more conceptual
than numerical. Given the number of particles a stochastic code is
often only mildly more complex (and it could even be simpler) at least
in the cases in which the noise is uncorrelated in time and acts on
one particle at a time. This means that the resulting code does not
require a longer running time than a deterministic one: a fact that
can also be seen by noting that a stochastic system can be regarded as
a deterministic system with more degrees of freedom (\ie the ones
needed to describe the random numbers generators that one has to use
and which, as it is well known, are simply suitably chaotic dynamical
systems themselves).

{}From the point of view of code writing this amounts at adding a few
more particles to the system.\footnotemark[8]\footnotetext[8]{For
instance if the stochastic thermostat is defined by requiring that
upon collision with the boundary a particle rebounds with a maxwellian
velocity distribution with dispersion (temperature) depending only on
the boundary point hit then one needs three Gaussian random number
generators, \ie essentially three more degrees of freedom.}

\kern3mm\0($\bullet$) It is by no means clear that by using mechanical
thermostats one can obtain physically realistic models nor, assuming
that the stochastic dynamic models are more realistic, models behaving in as
complex a way as the stochastic ones (typically consisting in boundary
collision laws in which the particles emerge with a maxwellian
velocity distribution with suitable variance, \ie suitable
temperature). Understandably the matter is controversial but quite a
few researchers think that this has been positively demonstrated by
large amounts of work done in the last thirty years.

Since stochastic models are just models with more degrees of freedom
it is tautological that there is equivalence between all possible
stochastic models and all deterministic ones. The real question is
whether the rather simple deterministic thermostats models that have
been used are able to simulate accurately the stochastic models
believed to be more realistic.

\kern3mm\0($\bullet$) In my view it is likely that vast classes of
thermostats, deterministic and stochastic as well, are equivalent in
the sense that they produce motions which although very different when
compared at equal initial conditions and at each time have,
nevertheless, the same statistical properties\Onlinecite{Ga00,Ga02}. And in
my opinion there is already evidence that it is indeed possible to
simulate the same system with simple deterministic thermostats or with
some corresponding stochastic ones.

Here the equivalence is intended in a sense that is familiar in the
theory of equilibrium ensembles: if one fixes suitably certain
parameters then ensembles (\ie time invariant probability
distributions in phase space) that are apparently very different (\eg
microcanonical and canonical) give, nevertheless, the same statistical
properties to vast ({\it not all}) classes of observables. If one
fixes the energy $U$ and the volume $V$ in a microcanonical ensemble or
the inverse temperature $\b$ and the volume $V$ in the canonical ensemble
then one obtains that local observables have the same statistical
distribution in the two cases provided the value of $\b$ is chosen
such that the average canonical energy is precisely $U$.

One among the most striking examples of such equivalence
(She,Jackson)\Onlinecite{SJ93} is the equivalence between the dissipative
Navier Stokes fluid and the Euler fluid in which the energy content of
each shell of wave numbers is fixed (via Gauss' least constraint
principle) to be equal to the value that Kolmogorov's theory predicts
to be the energy content of each shell at a given (large) Reynolds
number. Here one compares two very different mechanical thermostats.
A more general view on the equivalence between different thermostats
has been developed since. In fact many instances in which Physicists
say that ``an approximation is reasonable'' really correspond to
equivalence statements about certain properties of different theories
(and in the best cases the statements can be translated into proper
mathematical conjectures).

\kern3mm\0($\bullet$) Coming therefore to consider more closely mechanical
thermostat models the phase space contraction has turned out in many
cases to be an interesting quantity often intepretable as the ratio
between the work done by the thermostats on the systems and some
kinetic energy average: this led since the beginning to identify the
phase space contraction rate with the entropy creation rate. The above
``philosophical'' considerations have been developed to give some
background interpretation to the vast phenomenology generated by the
new electronic machines used as tools to investigate complex systems
evolutions.

A collision between the previously held views, egregiously summarized
by the book of De Groot-Mazur\Onlinecite{DGM84}, based on {\it continuum
mechanics} and the new approaches based on {\it transistors, chips and
dynamical systems theory} ensued: often showing that the two
communities give the impression of not really meditating on each other
arguments.

\kern3mm\0($\bullet$) Setting aside controversies it is interesting that the
mechanical thermostats approach has nevertheless led to a new
perspective and to a few new results. Here I mention the {\it
fluctuation relation}: the phase space contraction $\s(x(t),\dot
x(t))$ which in various models has interesting physical meaning (like
being related to conductivity or viscosity) is a fluctuating quantity
as time goes on. Its average value in a time interval of size $\t$
divided by its infinite time average in the future $\media\s$ is a
quantity $p$ that still fluctuates. Of course it fluctuates less and
less the larger is $\t$ and its probability distribution (easily
analyzable by observing it for a long time and by dividing the time
into intervals of size $\t$ and forming a histogram of the values thus
observed) is expected on rather general grounds to be proportional to
$e^{\t\z(p)}$ for $\t$ large with $\z(p)$ being a function with a
maximum at $p=1$ (\ie at the average, hence most probable, value of
$p$) {\it provided $\media{\s}>0$}, \ie provided there is dissipation
in the system and the system is, therefore, out of equilibrium.

If the dynamical equations are {\it reversible}, \ie if there is an
isometry $I$ of phase space which anticommutes with the time evolution
$(x,\dot x)\to S_t(x,\dot x)\equiv(x(t),\dot x(t))$ in the sense that
$I S_t=S_{-t}I$ and furthermore $I^2=1$ then, provided the
system motion is ``very chaotic'', it follows that

$$\z(-p)=\z(p)-p\media{\s}$$
This is a {\it parameter free} symmetry relation that was discovered in a
numerical experiment (Evans, Cohen, Morriss)\Onlinecite{ECM93} and which has been
checked in many cases. By very chaotic one means that the motion of
the system can be assimilated to that of a suitable Anosov flow whose
trajectories fill densely phase space (transitive Anosov flow).

\kern3mm\0($\bullet$) Indeed for transitive reversible Anosov systems
the above relation holds as a theorem\Onlinecite{GC95}. Since models of
physical systems are {\it not} Anosov systems from a strict
mathematical point of view the above relation cannot be applied, not
even to cases in which the model is reversible and the trajectories
are dense on the allowed phase space: the chaotic hypothesis says that
the fact that the system is not mathematically an Anosov system is not
relevant for physical observations, in most cases. This is similar to
the statement that in equilibrium systems the lack of ergodicity of
motions is irrelevant in most cases and averages can be computed by
assuming ergodicity (\ie by using the microcanonical distribution).

If this is correct the above relation should hold: a non trivial fact
to check due to the difficulty of observing such large fluctuations
{\it and} to the lack of free parameters to fit the data, once they have
been laboriously obtained.

\kern3mm\0($\bullet$) When the forcing of the system is let to $0$ the above
relation degenerates: not only $\media{\s}\to0$ but also $p$ itself
becomes ill defined as its definition involves division by $\media\s$.
Nevertheless by extracting the leading behavior of both sides the
fluctuation relation leads to relations between average values of
derivatives of dynamical quantities with respect to the intensity of
the forcing, {\it evaluated at zero forcing}, and such relations can be
intepreted as Onsager reciprocity relations and Green--Kubo
expressions for suitably defined transport coefficients\Onlinecite{Ga02}.

\kern3mm\0($\bullet$) Clearly a reversibility assumption on thermostats is a
strong assumption and so is the chaotic hypothesis. Nevertheless the
results are interesting and they seem to be among the few that can be
obtained in a field which is well known for its imperviousness. The
philosophical framework developed in Sections 1,2 helps keeping a
unified view on a subject that is being developed although, strictly
speaking, one could dispense with the philosophical view and
concentrate on obtaining results that can be drily stated without
appealing to entropy, entropy creation, thermostats {\it etc}.

\kern3mm\0($\bullet$) And one can go beyond various assumptions via
the use of equivalence conjectures between different thermotats: for
instance Drude's thermostat model which strictly speaking is not
reversible is conjectured to be equivalent to a Gaussian thermostat
which is reversible. The Navier Stokes equation for incompressible
fluids, clearly irreversible, is conjectured to be equivalent to a
similar reversible equation\Onlinecite{Ga02} as the quoted
experiment\Onlinecite{SJ93} shows and as other successive experiments seem
to confirm\Onlinecite{GRS02}. The research along the just mentioned lines
seems to go quite far and to lead not only to new perspectives but
also to new results or confirmations (\ie non contradictions) of the
general views in Sections 1,2. Doubts about the whole approach can be
legitimately raised, and have been raised, on the grounds that the
results are too few and too meager to be really interesting: for
instance one can hold against their consideration that they are not
even sufficient to give some hint at a derivation of ``elementary''
relations like Fourier's law or Ohm's law. One can only say that time
is not yet ripe to see whether the new methods and ideas lead really
anywhere or at least to a better understanding of some of the problems
that also the old ones have not been able to tackle, so far, (like the
heat conduction laws or the electric conduction laws) in spite of
intense research efforts.%

\kern3mm
\0{\it The above comments have been stimulated by continuous heated
discussions held at Rutgers University in the course of the last few
years: involving among many others S.Goldstein, J.Lebowitz, D.Ruelle.}

\kern3mm
\0e-mail {\tt gallavotti@roma1.infn.it}\\
Piscataway, NJ, november 2002

\end{section}
\bibliography{ELA} \bibliographystyle{unsrt}

%%%%%\revtex
\end{document}